\documentclass[aps,prd,eqsecnum,preprint,tightenlines,nofootinbib,showpacs]{revtex4-1}
\usepackage{graphicx,amsmath,latexsym}

\def\bea{\begin{eqnarray}}
\def\eea{\end{eqnarray}}
\def\be{\begin{equation}}
\def\ee{\end{equation}}
\newcommand{\ub}[1]{\underline{#1}}

\begin{document}

\title{Light-front analysis of the Casimir effect}

\author{Sophia S. Chabysheva}
\author{John R. Hiller}
\affiliation{Department of Physics \\
University of Minnesota-Duluth \\
Duluth, Minnesota 55812}

\date{\today}

\begin{abstract}
The Casimir force between conducting plates at rest in an inertial frame
is usually computed in equal-time quantization, the natural choice
for the given boundary conditions.  We show that the well-known result
obtained in this way can also be obtained in light-front quantization.
This differs from a light-front
analysis where the plates are at ``rest'' in an infinite momentum frame,
rather than an inertial frame; in that case, as shown by Lenz and Steinbacher,
the result does not agree with the standard result.  As is usually done, 
the analysis is simplified
by working with a scalar field and periodic boundary conditions, in place of
the complexity of quantum electrodynamics.  The two key ingredients
are a careful implementation of the boundary conditions, following the work of
Almeida {\em et al.} on oblique light-front coordinates, and computation
of the ordinary energy density, rather than the light-front energy density.
The analysis demonstrates that the physics of the effect is independent
of the coordinate choice, as it must be.
\end{abstract}

\pacs{03.70.+k, 11.10.Ef} 


\maketitle

\section{Introduction}
\label{sec:Introduction}

The Casimir effect~\cite{Casimir,ItzyksonZuber} is the existence of a force between
conducting plates due to the exclusion of vacuum modes by boundary conditions at the plates.
The vacuum energy density between the plates differs from the free density
and defines an effective potential energy for the plates that varies with
the plate separation.  The energy density is computed by summing over
the allowed modes, and the variation of the effective potential yields
the force.  To simplify these calculations, the physical boundary conditions
at the conducting surfaces can be replaced by periodic boundary conditions,
and the photon field replaced by a massless scalar field.

Because the plates are at rest in an inertial frame, the natural
choice of coordinates for the analysis is the standard set of equal-time
coordinates, rather than Dirac's light-front coordinates~\cite{Dirac,DLCQreview}.
Nevertheless, attempts at analysis in light-front quantization
have been made~\cite{Lenz,Almeida}.  In Ref.~\cite{Lenz}, the analysis
considered the light-front analog of spatial periodicity, with
boundary conditions periodic in $x^-\equiv t-z$ rather than $z$.
However, this corresponds to plates moving with the speed of
light, which cannot be realized experimentally, and, in any case,
the calculation did not lead to a well-defined Casimir force.
In Ref.~\cite{Almeida}, the coordinate choice was modified away
from proper light-front coordinates in such a way as to avoid the
difficulties encountered by Lenz and Steinbacher.  In their analysis,
Almeida {\em et al}.\ arrived at suitable boundary conditions,
conditions that mix time and space coordinates.  This provided
the motivation for our approach to a truly light-front analysis,
where we require periodicity in $z$ not $x^-$.  Such a choice
is not ``natural'' for light-front coordinates because it
mixes $x^-$ with light-front time $x^+\equiv t+z$.  However,
a calculation must be dictated by the physics, not the coordinates.

A careful choice of boundary conditions is not the entire story.
To complete a calculation of the Casimir effect in
light-front coordinates, we must calculate the true vacuum energy,
not the light-front energy $p^-\equiv E-p_z$.  The ordinary energy $E$
is what determines the effective potential, which in turn determines
the Casimir force.  That the physics of a system is determined by $E$
was seen in a light-front variational analysis of $\phi^4$ theory by 
Harindranath and Vary~\cite{Vary}
and in finite-temperature calculations by Elser and Kalloniotis~\cite{Elser}.
In the latter case, the important point is that a partition function should
be computed for contact with a heat bath at rest in an inertial
frame; if the light-front energy $p^-$ is used instead of $E$,
the heat bath must be interpreted as moving with the speed of light,
an unphysical situation.\footnote{For further discussion of this point,
see Ref.~\protect\cite{SDLCQ}.}

\begin{figure}[ht]
\vspace{0.2in}
\centerline{\includegraphics[width=8cm]{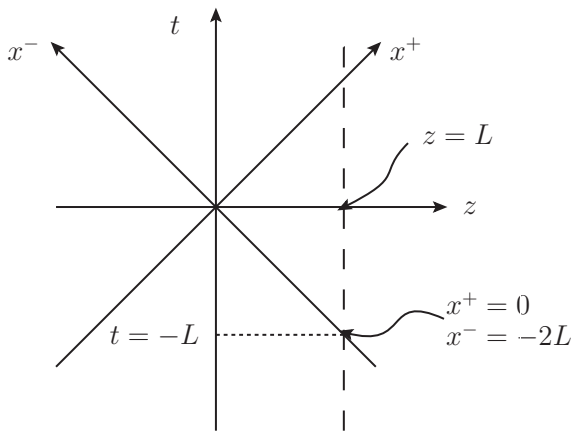}}
\caption{\label{fig:Fig1} Light-front coordinates for the
longitudinal case.  The periodicity is in $z$, from 0 to $L$.
The line for $z=L$ intersects the $x^-$ axis at $x^-\equiv t-z=-2L$.
}
\end{figure}

Our definition of light-front coordinates is illustrated in Fig.~\ref{fig:Fig1}.
%
%
In addition, we define transverse spatial coordinates $\vec x_\perp=(x,y)$
and the light-front momentum $\ub{p}=(p^+\equiv E+p_z,\vec{p}_\perp)$.
The scalar product of four-momentum and coordinates is then given by
$p\cdot x=\frac12(p^+x^- + p^- x^+)-\vec{p}_\perp\cdot\vec{x}_\perp$, and
the mass-shell condition $p^2=m^2$ yields $p^-=(p_\perp^2+m^2)/p^+$.
We will make some use of these relations in the following sections.

The remainder of the paper contains our analysis of the Casimir
effect for plates perpendicular to the $z$ axis, in Sec.~\ref{sec:longitudinal},
and for plates perpendicular to the $x$ axis, in Sec.~\ref{sec:transverse}.
The latter case, where periodicity in the spatial coordinate is the
same for both equal-time and light-front coordinates, is considered
in order to check that our approach has not destroyed the agreement
between coordinate systems already obtained by Lenz and Steinbacher~\cite{Lenz}.
A summary is given in Sec.~\ref{sec:summary}.

\section{Longitudinal case}
\label{sec:longitudinal}

We first consider plates separated in the $z$ direction, as shown
in Fig.~\ref{fig:Fig1}.  One plate is at $z=0$, and the other at $z=L$.
The standard result for the expectation value of the energy
density is a sum over zero-point energies~\cite{ItzyksonZuber}
\be \label{eq:stdresult}
\langle{\cal H}\rangle=\frac{1}{2L}\sum_{n=-\infty}^\infty \int \frac{d^2p_\perp}{(2\pi)^2}E_n,
\ee
with
\be \label{eq:En}
E_n=\sqrt{p_\perp^2+\left(\frac{2\pi n}{L}\right)^2}.
\ee
The sum is typically regulated by a heat-bath factor\footnote{In
light-front quantization, the same factor should be used, since the
system should be in contact with a heat bath at rest in an inertial frame.}
$e^{-\Lambda E_n}$.  The sum then yields
\be
\langle{\cal H}\rangle=\frac{3}{2\pi^2 \Lambda^4}-\frac{\pi^2}{90 L^4}.
\ee
The second term provides the effective potential, independent of the
regulator, and determines the Casimir force.

To simulate the Casimir effect, we impose periodic boundary conditions
on a neutral massless scalar field and compute the vacuum energy density.
The mode expansion for the scalar field is
\be \label{eq:mode}
\phi=\int \frac{d\ub{p}}{\sqrt{16\pi^3 p^+}}
   \left\{ a(\ub{p})e^{-ip\cdot x} + a^\dagger(\ub{p})e^{ip\cdot x}\right\},
\ee
with the modes quantized such that 
\be
[a(\ub{p}),a^\dagger(\ub{p}')]=\delta(\ub{p}-\ub{p}').
\ee
The periodicity imposed is $\phi(z+L)=\phi(z)$; in light-front
coordinates, this is
\be
\phi(x^++L,x^--L,\vec{x}_\perp)=\phi(x^+,x^-,\vec{x}_\perp).
\ee
This implies $-p^+L/2+p^-L/2=2\pi n$ or
\be
\frac{p_\perp^2}{p^+}-p^+=\frac{4\pi}{L}n,
\ee
with $n$ any integer between $-\infty$ and $\infty$.
The positive solution of this constraint is
\be
p_n^+\equiv \frac{2\pi}{L}n+\sqrt{\left(\frac{2\pi}{L}n\right)^2+p_\perp^2}.
\ee
Then $n=-\infty$ corresponds to $p^+=0$, and $n=\infty$ to $p^+=\infty$.

The mode expansion of the field is restricted to a 
discrete sum for the longitudinal modes.  We define discrete
annihilation operators
\be
a_n(\vec{p}_\perp)=\sqrt{\left|\frac{dp^+}{dn}\right|}\;a(p_n^+,\vec p_\perp),
\ee
for which the commutation relation becomes
\be
[a_n(\vec{p}_\perp),a_{n'}^\dagger((\vec{p}_\perp)^{\,\prime})]
    =\delta_{nn'}\delta(\vec{p}_\perp-\vec{p}_\perp^{\,\prime}),
\ee
and change the integration over $p^+$ to a sum over $n$
\be
\int dp^+ =\int \frac{dp^+}{dn}dn\rightarrow \sum_n \frac{dp^+}{dn},
\ee
where $\frac{dp^+}{dn}=\frac{2\pi}{L} \frac{p_n^+}{E_n}$.
Substitution then gives
\bea
\phi(x^+=0)&=&\frac{1}{\sqrt{2L}}\sum_n \int \frac{d^2p_\perp}{2\pi\sqrt{E_n}}
   \left\{ a_n(\vec{p}_\perp)e^{-ip_n^+x^-/2+i\vec{p}_\perp\cdot\vec{x}_\perp} \right.  \\
  && \left. \rule{1.3in}{0mm}
     + a_n^\dagger(\vec{p}_\perp)e^{ip_n^+x^-/2-i\vec{p}_\perp\cdot\vec{x}_\perp}\right\},
     \nonumber
\eea
where the leading $\frac{1}{\sqrt{2L}}$ factor is consistent with the
normalization of the discrete basis functions
$e^{-ip_n^+x^-/2+i\vec{p}_\perp\cdot\vec{x}_\perp}$ on the interval
$-2L<x^-<0$.

For the free scalar, the light-front energy and longitudinal momentum
densities are ${\cal H}^-=\frac12|\vec\partial_\perp\phi|^2$
and ${\cal H}^+=2|\partial_-\phi|^2$.  Their vacuum expectation
values are
\bea
\langle0|{\cal H}^-|0\rangle&=&\frac{1}{4L}\sum_{n,n'}\int 
      \frac{d^2p_\perp d^2p'_\perp}{(2\pi)^2\sqrt{E_n E_{n'}}}
      \vec{p}_\perp\cdot\vec{p}_\perp^{\,\prime}
      \langle0|a_n(\vec{p}_\perp)a_{n'}^\dagger(\vec{p}_\perp^{\,\prime})|0\rangle \nonumber \\
      &=&\frac{1}{4L}\sum_n \int \frac{d^2p_\perp}{(2\pi)^2 E_n}p_\perp^2
\eea
and
\bea
\langle0|{\cal H}^+|0\rangle&=&\frac{2}{2L}\sum_{n,n'}\int 
      \frac{d^2p_\perp d^2p'_\perp}{(2\pi)^2\sqrt{E_n E_{n'}}} \frac{p_n^+ p_{n'}^+}{4}
      \langle0|a_n(\vec{p}_\perp)a_{n'}^\dagger(\vec{p}_\perp^{\,\prime})|0\rangle \nonumber \\
      &=&\frac{1}{4L}\sum_n \int \frac{d^2p_\perp}{(2\pi)^2 E_n} (p_n^+)^2.
\eea
These yield an energy density
\bea
{\cal E}_{\rm LF}&\equiv&\frac12(\langle0|{\cal H}^-|0\rangle+\langle0|{\cal H}^+|0\rangle) \\
&=& \frac{1}{8L}\sum_n \int \frac{d^2p_\perp}{(2\pi)^2 E_n} (2E_n^2+2\frac{2\pi}{L}nE_n) 
\eea
relative to light-front coordinates.  The second term is zero, because it is proportional
to $\sum_{n=-\infty}^\infty n=0$.  We then obtain
\be
{\cal E}_{\rm LF}=\frac{1}{4L}\sum_n\int \frac{d^2p_\perp}{(2\pi)^2}E_n.
\ee
However, we still need to relate this to the energy density relative to 
equal-time coordinates, which we denote simply by ${\cal E}$.

Integration over a finite volume between the plates yields
\be
{\cal E}=\frac{1}{LL_\perp^2}\int_{-2L}^0 dx^- \int_0^{L_\perp} d^2x_\perp {\cal E}_{\rm LF}.
\ee
A change of variable from $x^-$ to $z=(x^+ + x^-)/2$ at fixed $x^+$ simplifies this to
\be
{\cal E}=\frac{1}{LL_\perp^2}\int_0^L 2dx^- \int_0^{L_\perp} d^2x_\perp {\cal E}_{\rm LF}=2{\cal E}_{\rm LF}.
\ee
Thus, the energy density is
\be
{\cal E}=\frac{1}{2L}\sum_n\int \frac{d^2p_\perp}{(2\pi)^2}E_n,
\ee
which matches exactly the standard result (\ref{eq:stdresult}).  When properly regulated,
the sum can be performed to extract the regulator-independent piece
and the force calculated from the derivative with respect to the separation.

\section{Transverse case}
\label{sec:transverse}

The transverse case is less problematic.  In fact, a direct
implementation of light-front coordinates, without any of the
considerations made here, does yield the correct result~\cite{Lenz}.
Therefore, there could be concern that the additional steps that
we have introduced will somehow destroy this agreement.  However,
this does not happen, as we show in this section.

Without loss of generality, let the periodicity be in the $x$ direction,
so that we require $\phi(x^+,x^-,x+L_\perp,y)=\phi(x^+,x^-,x,y)$.
This is satisfied if $p_x$ is restricted to the discrete
values $p_n\equiv 2\pi n/L_\perp$.  We define discrete
annihilation operators
\be
a_n(p^+,p_y)=\sqrt{\frac{2\pi}{L}}a(p^+,p_n,p_y),
\ee
with the commutation relation
\be
{[}a_n(p^+,p_y),a_{n'}^\dagger(p^{\prime +},p'_y]
=\delta_{nn'} \delta(p^+-p^{\prime +}) \delta(p_y-p'_y).
\ee
The scalar field is then
\bea
\phi(x^+=0)&=&\frac{1}{\sqrt{L_\perp}}\sum_n \int \frac{dp^+ dp_y}{\sqrt{8\pi^2 p^+}}
   \left\{ a_n(p^+,p_y)e^{-ip^+x^-/2+ip_n x+ip_y y} \right.  \\
  && \left. \rule{1.3in}{0mm}
     + a_n^\dagger(p^+,p_y)e^{ip^+x^-/2-ip_n x -i p_y y}\right\}.
     \nonumber
\eea
The leading factor is consistent with the normalization of the wave functions
$e^{-ip^+x^-/2+ip_n x+ip_y y}$ on the interval $0<x<L_\perp$.

The energy and longitudinal momentum densities are
\bea
\langle0|{\cal H}^-|0\rangle&=&\frac{1}{2L_\perp}\sum_{nn'}
  \int\frac{dp^+dp_ydp^{\prime+}dp'_y}{8\pi^2\sqrt{p^+p^{\prime+}}}
  (p_n p_{n'}+p_y p'_y)  \nonumber \\
  && \rule{1.7in}{0mm} \times
    \langle0|a_n(p^+,p_y)a_{n'}^\dagger(p^{\prime+},p'_y)|0\rangle \nonumber \\
  &=&\frac{1}{2L_\perp}\sum_n \int\frac{dp^+ dp_y}{8\pi^2} \frac{p_n^2+p_y^2}{p^+}
\eea
and
\bea
\langle0|{\cal H}^+|0\rangle&=&\frac{2}{L_\perp}\sum_{nn'}
   \int\frac{dp^+dp_ydp^{\prime+}dp'_y}{8\pi^2\sqrt{p^+p^{\prime+}}}
   \frac{p^+ p^{\prime +}}{4}\langle0|a_n(p^+,p_y)a_{n'}^\dagger(p^{\prime+},p'_y)|0\rangle \nonumber \\
     &=&\frac{1}{2L_\perp}\sum_n \int\frac{dp^+ dp_y}{8\pi^2} p^+.
\eea
Averaged together, these determine ${\cal E}_{\rm LF}$ to be
\be
{\cal E}_{\rm LF}=\frac{1}{2L_\perp}\sum_n \int \frac{dp^- dp^+ dp_y}{8\pi^2}
   \frac{p^- + p^+}{2} \delta\left(p^--\frac{p_n^2+p_y^2}{p^+}\right).
\ee
The delta function is equivalent to the mass-shell condition:
\be
\delta\left(p^--\frac{p_n^2+p_y^2}{p^+}\right)=p^+\delta(p^2)=p^+\delta(E^2-E_n^2),
\ee
with $E_n=\sqrt{\left(\frac{2\pi}{L_\perp}n\right)^2 +p_z^2+p_y^2}$,
and facilitates a conversion to integration over the equal-time variables
$E=(p^+ +p^-)/2$ and $p_z=(p^+-p^-)/2$.  The conversion yields
\be
{\cal E}_{\rm LF}=\frac{1}{2L_\perp} \sum_n \int \frac{2dE dp_z dp_y}{8\pi^2}
   E(E+p_z)\frac{1}{2E_n}\delta(E-E_n).
\ee
The $E$ integral can be done immediately, with use of the delta function.
The contribution from the $p_z$ term is zero, because that part of the $p_z$
integral is trivially odd.\footnote{This explains why our approach yields the
same result as a calculation of the light-front energy density alone in the 
transverse case; the difference is just contributions proportional to $p_z$,
which integrate to zero.}  This leaves an energy density relative to light-front
coordinates of
\be
{\cal E}_{\rm LF}=\frac{1}{4L_\perp}\sum_n\int\frac{dp_z dp_y}{(2\pi)^2}E_n.
\ee
The transformation to the energy density relative to equal-time coordinates
is, as before, just multiplication by two.  Therefore, we obtain in the
transverse case
\be
{\cal E}=\frac{1}{2L_\perp}\sum_n\int\frac{dp_z dp_y}{(2\pi)^2} E_n,
\ee
which matches the usual equal-time result and is of the same form
as in the longitudinal case.

\section{Summary}
\label{sec:summary}

By a physical choice of boundary conditions and vacuum energy, we have
computed in light-front coordinates the vacuum energy density appropriate for
the Casimir effect~\cite{Casimir} and obtained the standard result 
(\ref{eq:stdresult}).  Unlike previous attempts~\cite{Lenz,Almeida},
we have invoked the physics of plates at rest in an inertial frame
and have not resorted to alteration away from true light-front
coordinates.  Keeping the plates at rest is not natural in light-front
coordinates, but is physically correct.  Similarly, the vacuum energy
is computed as the standard equal-time energy, which is the correct
input to the calculation of a Casimir force.  Again, this is not
the ``natural'' choice in light-front coordinates, where one would
usually calculate the light-front energy.  In other words, by being
careful to calculate the same physical quantity, we have obtained
the standard result, though with use of a different coordinate system.

Clearly, light-front coordinates are not the preferred system for
this calculation.  However, there are many situations in nonperturbative
field-theoretic calculations where light-front coordinates are much
superior~\cite{DLCQreview}.  In particular, Fock-state expansions are
well-defined and the associated wave functions are boost invariant,
making the calculation of observables relatively straightforward.
That something such as the Casimir effect, which is much less natural
for light-front coordinates, can also be calculated correctly provides
additional confidence in the usefulness of the approach.

\acknowledgments
This work was supported in part by the U.S. Department of Energy
through Contract No.\ DE-FG02-98ER41087.


\end{document}